\documentclass[fleqn,10pt]{wlscirep}

\title{Identification and acceptation of macroscopic magnetism energy levels results in better understanding of the linkages between the traditional theory with quantum electrodynamics and revelation of the limited validity of the Faraday's Law.}
\author[1,*]{Pavol Ivana}
\author[2]{Marika Ivanova}
\affil[1]{Supratech s.r.o. scientific-technical laboratory, Mokra 314, Zlin, 76001, Czech Republic}
\affil[2]{University of Bergen Faculty of Mathematics and Natural Sciences, PO Box 7803, Bergen, 5020, Norway}
\affil[*]{info@supratech.org}

\begin{abstract}
The construction of the Pure Homopolar Generator (PHG) reveals the physical problem of Maxwell's equations. In this article, we present the result of the research, which was directed at gaining electromotive voltage in a theoretically pure way. We designed and built a brushless generator that simulates a homogenised magnetic field and should theoretically be usable without semiconductors and electronic components, because the motion of the conductor towards the induction vector is in the relation $\vec v \perp \vec B$. The generator is equipped with superconductive shielding, which ensures disruption of the theoretical balance of electromotive voltage generation – in the Faraday homopolar generator, the imbalance of the electromotive voltage is secured by fixing the disk to the reference set of rotating magnets. However, the solution of the technical problem initiated a theoretical problem. The result of the experiment suggests that the current concept of electrodynamics, based on magnetic flux using relativistic principles, is Euclidean, idealised and misleading. If we continue to persist in the unconditional correctness of Maxwell's concept, we would have to admit that the inhomogeneous field can be screened out from the perspective of any external reference system, but that for a homogeneous magnetic field such a reference system would not exist. A good explanation of the inconsonance with theoretical expectations gives us the introduction of energy levels of the magnetic field, which are measurable and are probably a macroscopic manifestation of the summarisation of levels from elementary particle environments. Part of this article is an analysis which shows that current electrodynamics use a simplified view of the vector of induction $\vec B$. We reveal that the collinearity $\vec v \parallel \vec B$, which implies $\vec E = 0$ is a special case of a more general, topological assessmen of the properties of the set of magnetic field vectors. This unavoidably leads to a narrowing of Faraday's Law, which improves the experimental prediction and paradoxically reveals considerable technical potential. Both concepts can coexist in practice, with a wide range of value matches. 
\end{abstract}
\begin{document}

\flushbottom
\maketitle% * 
%
%  Click the title above to edit the author information and abstract
%
\thispagestyle{empty}
\begin {multicols}{2}
%\twocolumn

\section*{Introduction}
2015 was the year of the 150th anniversary of the genesis of classic Maxwell equations. We therefore decided to confront the academic public with the results of our research and analysis. We have reached conclusion that it is necessary to acquaint the wider professional community with the negative results of the experiment carried out on a superconductor shielded PHG and to familiarise them with the design of a concept that can explain this negative result. The results point to a contradiction of the physical nature of what is called Faraday’s Law, describing the formation of electromotive voltage (EMF) as the unconditional consequence of the time change of magnetic flux.
The result shows that Faraday's Law is conceived too generally and, on the other hand, the properties of the vector of magnetic induction $\vec B$ are too special. Faraday's Law is not able to explain the non-functioning of a shielded PHG; it represents a weak condition for the creation of electromagnetic induction. In this article, we analytically propose a dual theory, based on a more complex interaction of the conductor with the magnetic field. We remove the myth of modern electrodynamics that the solving of Faraday’s homopolar generator, (FHG), could exist independently of brushes and electronic elements using a homogenised magnetic field.

\section*{Results}

\hspace{0,45cm}
{\bfseries Motivation and performed experiments}
  
In 2012, we made an attempt to revive industrial applications based on FHG. The drawback of FHG is the necessity of using brushes. Therefore, an experimental homopolar generator was designed to eliminate this drawback.  

Our proposal corresponds precisely to the contemporary theoretical ideas of FHG function. Mathematically it is described by the well-known equation formulated by James Clerk Maxwell in about 1865 \,\cite{maxw,hall}
\begin {eqnarray}  \label{U_m}
  %\mathcal U_m = 
  \mathcal E_m=
  - \frac{d \Phi}{d t}
   \ \ = \ \
   \oint\limits_l \vec E_m \cdot \ d \vec l = \int\limits_S rot\vec E_m\cdot d\vec S ,
\end {eqnarray}
where index \emph{m} means determination of origin: \emph{Maxwellian value}. Figures \ref{obr0}, \ref{obr1}, \ref{obr1a} and \ref{obr1b} show the method of brush elimination. In PHG, as shown in Fig. \ref{obr0}, the conductance path enters the isomagnetic radials of the magnetic field perpendicularly at two opposite, axially magnetised, synchronously rotating rings. The output of the conductivity path is provided by the central conductor passing through the hollow shaft as shown in \ref{obr1b}. Assuming that we provide superconductor shielding of a wire $\vec j$ of measuring circuit inlet, $S$, PHG function in Figure \ref{obr0} will theoretically be similar to the brush solution of FHG \cite{tes}. To achieve the correct function of such shielding, a YBaCuO crystal based superconductor was selected \cite{can}. Shielding has two aspects:
\begin{center}
\scalebox{0.325}{\includegraphics{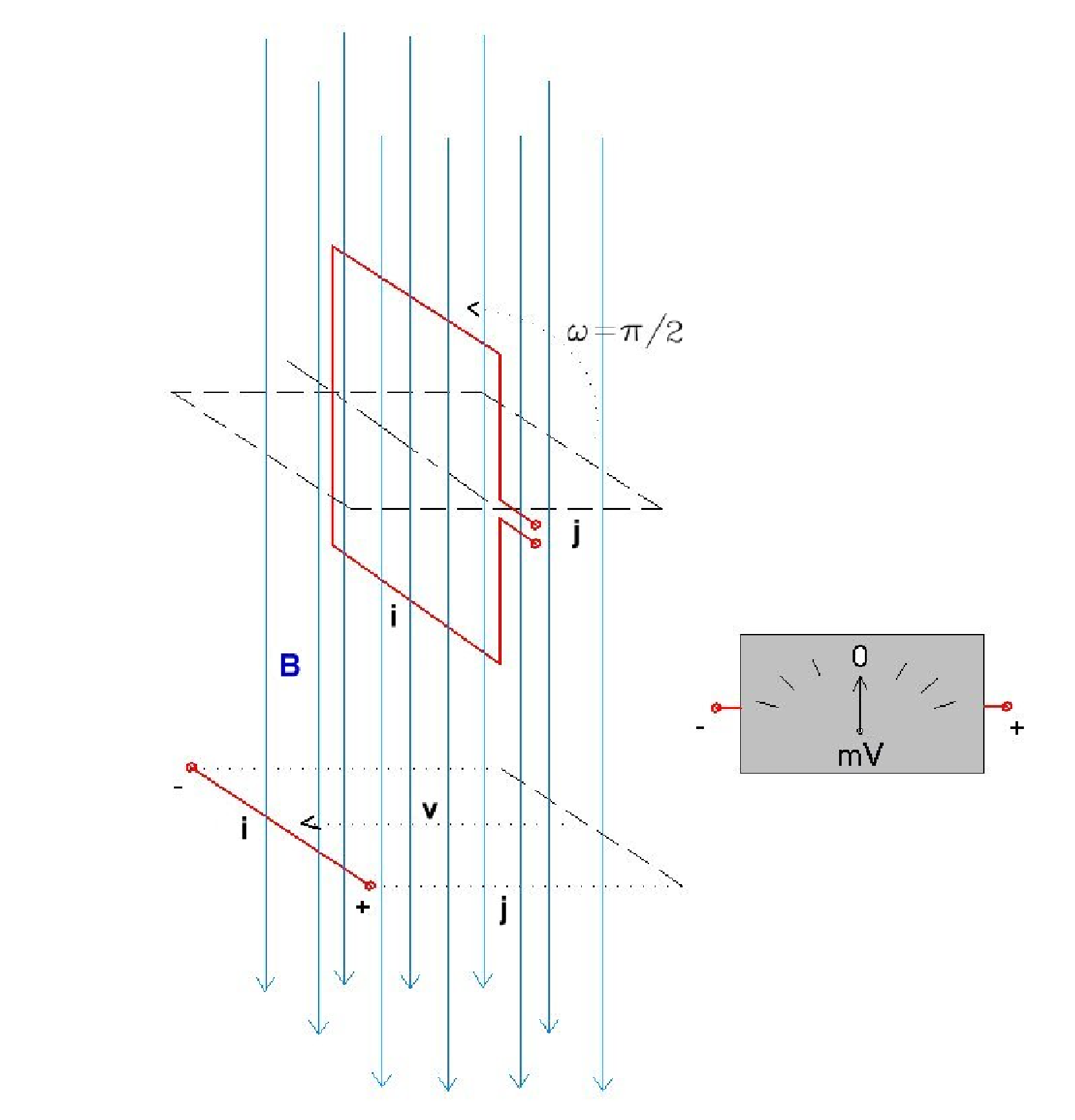}}\\
\captionof{figure}{\footnotesize{The Upper part of the image graphically illustrates the special case of a rectangular loop rotation of angle $\pi/2$, which causes a change of homogenous magnetic flux $\Phi=\vec S.\vec B$ from maximum to minimum, and which is equivalent to the movement of the conductor of length $\vec i$ in the path $\vec j$ at the bottom of the image. In both cases, the same mean EMF $\mathcal E_m$ should be induced.}}
\label{obrz}
\end{center}
\begin{itemize}
  \item It provides a comparison of activity between PHG and FHG from the point of the Lorentz force creation \cite{lore} - the shielded part of the wire will not interact with the external moving field just as the case is with the inner part of the FHG disc. Deformation of the surrounding field adjacent to the superconductor shielding has no impact on the final stability of the field with respect to the closed loop $l=i+j+k+i^{\textbf,}+k^{\textbf,}$. 
  \item It ensures theoretical induction imbalance between positive and negative contributions of EMF on the loop of length $l$ according to the line integral in equation \eqref{U_m}. This will allow assessment of the induction occurrence in PHG from the viewpoint of equations derived from the continuity equation. 
\end{itemize}          
In theory, the function of the brushless model of Faraday's generator should be equivalent to the brush solution. Current theory is based on the fact that the cause of induction can be explained, in terms of general geometry, using Stokes' Theorem. This theorem converts the line integral of electrical intensity $\vec E_m$ over the length of the conductive coil to a surface rotation integral $rot\vec E_m$ over the surface delimited by the coil 
in accordance with \eqref{U_m}. The emergence of electric intensity $\vec E_m$ should therefore be conditioned by the direction and velocity $\vec v$ of the elementary lengths of winding in a homogeneous magnetic field of magnetic
\begin{center}
 \scalebox{0.48}{\includegraphics{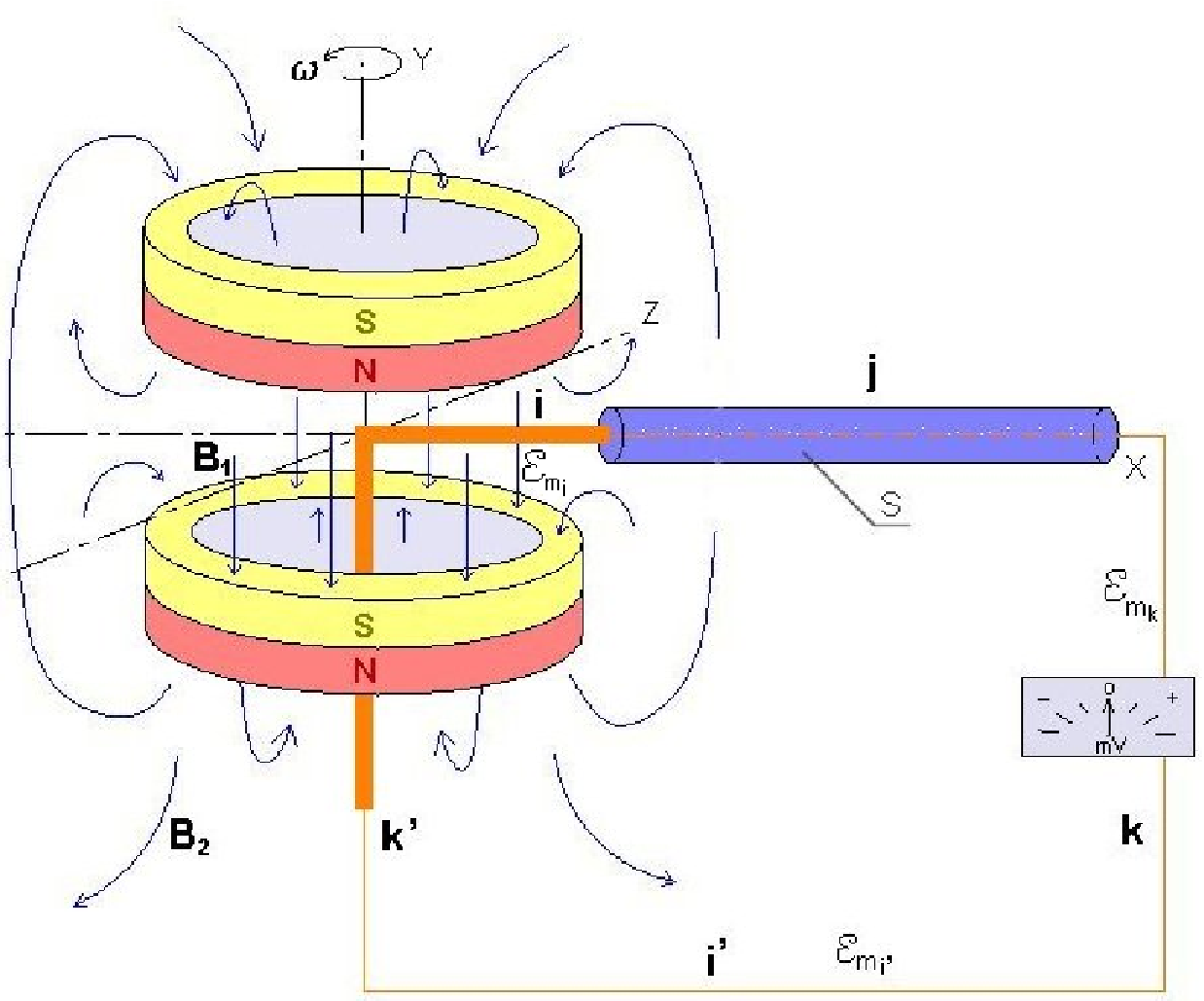}}\\
\captionof{figure}{\footnotesize{
Axonometric illustration of PHG function: Neodymium magnetic rings rotate at the angular velocity of $\omega$ with respect to a standing enclosed wire $l=i+j+k+i^{\textbf,}+k^{\textbf,}$. The scheme shows the superconductor shielding $S$, which prevents the external magnetic field from entering the wire segment $\vec j$.
}}
\label{obr0}
\end{center}    
induction $\vec B$, or equivalently by changing of magnetic flux $\Phi=\vec S.\vec B$ through the surface $\vec S$ encircled during time $\Delta t$. Ultimately, this situation can be simplified and represented as shown in Figure \ref{obrz}, where in the upper part, the loop performs axial rotation by a quarter of the period. This partial turn of one winding creates a relative change in area $\vec S=\vec i.\vec j  \rightarrow  0$  against the magnetic induction $\vec B$ during time $\Delta t$, which is encircled by the winding, and which the imaginary magnetic flux $\Phi$ of homogenous magnetic field passes. A mean $\mathcal E_m=\Phi/ \Delta t =\vec i .\vec j .\vec B/\Delta t=\vec i.\vec v. \vec B$ should thus be induced. \,According to this theorem, it is therefore equivalent to moving of the wire with length $\vec i$ along the path $\vec j$ with the speed $\vec v=\vec j/\Delta t$. So, even in the latter case, EMF $\mathcal E_m=\vec E_m.\vec i=\vec v.\vec B.\vec i$ \ should be induced with the same mean value as in the previous case. According to Figure \ref{obr0}, it is obvious that the conductor $\vec i$ at PHG performs a similar relative movement as in the bottom of Figure \ref{obrz} with the fact that this movement is circularly oriented.           
\begin{center}                                                   
\scalebox{0.42}{\includegraphics{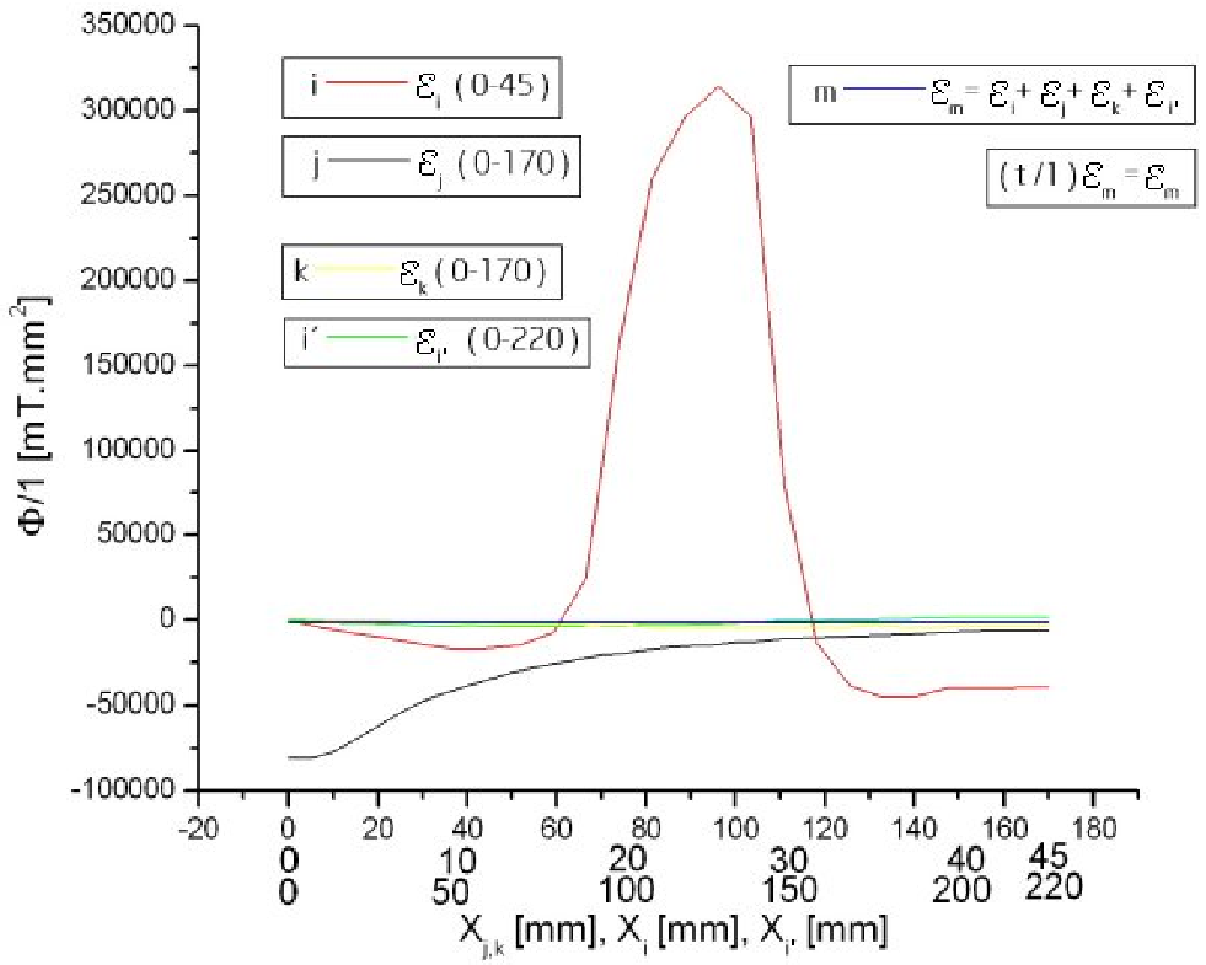}}\\
\captionof{figure}{\footnotesize{Chart of inductive flows per one turn that occur in closed PHG loop without shielding. The blue line represents the resulting sum of the $\mathcal E_{m}$ at one revolution per second. The red line represents the course of contributions $\mathcal E_{m_i}$ on the path $\vec i$ between the magnets. The black line represents the course of the largest external contributions ${\mathcal E}_{m_j}$ of the loop part $\vec j$, which is undesirable and must be shielded off. The yellow line represents the course of small contributions $\mathcal E_{m_{k}}$ of the loop part $\vec k$. The green line represents the residual course of contributions $\mathcal E_{m_{i^{\textbf,}}}$ of the part of the loop $\vec i^{\textbf,}$. The graph was created by the sum of measured values in accordance with the line integral in equation \eqref{U_m}}}
\label{PHG-gr}
\end{center}    
Now, let us analyse the application of current theory to PHG as shown in Fig. \ref{obr0} without shielding $S$: For the surface integral, we consider the plane $[x,y]$, which is parallel to the measuring loop; thus we obtain $\Phi = 0$ for surface $\vec S=(\vec i+\vec j) \times \vec k = \vec i^{\textbf,} \times \vec k^{\textbf,}$ delimited by this loop. The induction flux in the unshielded PHG will be constant, inducing no voltage. With respect to the line integral in \eqref{U_m}, the external magnetic flux lines are opposed to the internal flux lines by direction (vector orientation) and their impact on the wire loop is inverse. Due to Lorentz force, which - in theory - influences free electrons by the relative motion of arm $\vec i + \vec j, \vec k, \vec i^{\textbf,}$ of the closed loop $l$ (the segment $\vec k^{\textbf,}$ being neutral) with respect to the magnetic induction vectors $\vec B$, the positive and negative EMF contributions are theoretically in balance, in accordance with the assumption $div\vec B = 0$ and with the experimentally compiled chart shown in Fig. \ref{PHG-gr}: $\mathcal E_m = \int\limits^{l}_0 \vec E_m \cdot \ d \vec l = 0 $. Therefore, there is a theoretical as well as experimentally proven equivalence between the induction flux through the loop area and the assumed voltage induced in the loop,  as per \eqref{U_m}. Fig. \ref{PHG-gr} shows the results of measurements taken at positions close to the PHG magnets.	   

For the compilation of Chart \ref{PHG-gr}, the perpendicular components magnetic induction $\vec B$ with respect to wire $l$ was conveniently selected. When measuring using a 3D teslameter, the most convenient method was to read the perpendicular components value directly on the display. This can be achieved by aligning one of the teslameter coordinates with the wire axis, while the second coordinate is aligned with the velocity vector and the third one is used to create the chart as shown in Fig. \ref{PHG-gr} or \ref{PHG-gr2}, respectively.

To be certain about Maxwell's equations \eqref{U_m}, we need to test the induction by moving the wire at least within the homogeneous radials of the magnetic field. Let us focus on the occurrence of Lorentz force in PHG, which is theoretically purer and more interesting in technical terms, as shown in Fig. \ref{obr0}. Thus we ensure the induction imbalance between the positive and negative EMF contributions in the closed loop as shown in the chart in Fig. \ref{PHG-gr2}.  
\begin{center}
\scalebox{0.45}{\includegraphics{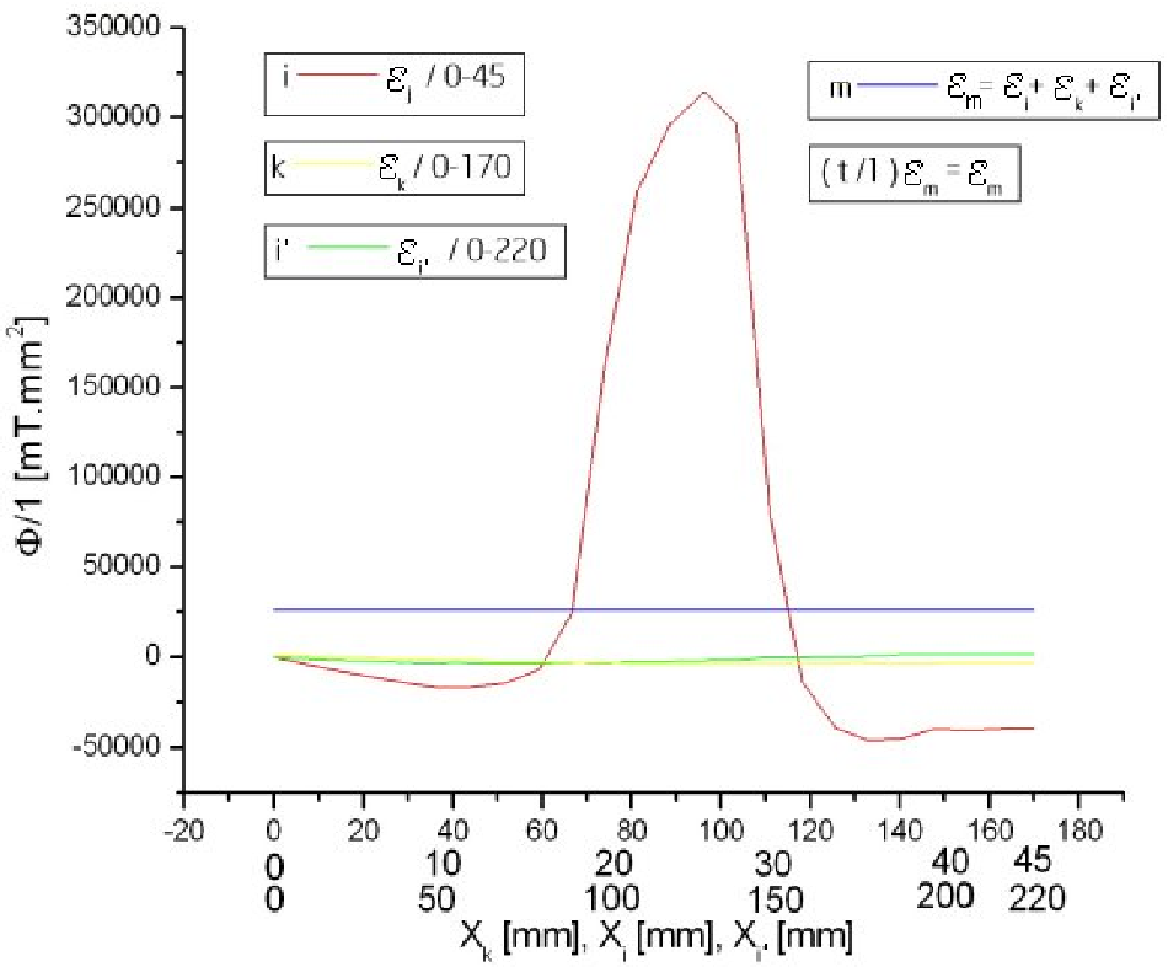}}\\
\captionof{figure}{\footnotesize{Similar to the previous graph of inductive flows per one revolution, which occur in the closed PHG loop at the shielding of the undesired return wire $\vec j$. The blue colour represents the resulting sum of the EMF $\mathcal E_{m}$ at one revolution per second, which can be assumed for a PHG adjusted in such a way.}}
\label{PHG-gr2}
\end{center}                       
The sectional magnetic flux cannot be used in PHG in this case, because by shielding the external part of the wire using the massive YBaCuO crystal-based superconductor shielding up to $100$ [mT] \cite{can}, we actually obtain an open circuit\footnote{The circuit is open in terms of the external field generated by the neodymium magnets.}. Similarly, the circuit is also open in these terms in FHG, where - on the contrary - the internal part of the conductance path represents a similarly shielded segment because the technical solution of this part cannot contribute to the EMF creation. The difference between PHG and FHG consists of the fact that in FHG, it is the external flux lines of magnetic induction (of a circularly homogenised magnetic field) are in relative motion with respect to the external standing frame with the brush and measuring loop wire. The internal part of FHG is standing, with respect to the magnet's flux lines, as both parts are fixed together\footnote{In technical practice and in media, we usually see the contrary description, which states that the Lorentz force is generated directly in the FHG disc\cite{radovic}.In terms of Maxwell's equations, this description is not complete but leads to identical mathematical results - this is caused by preserving the continuity for both internal and external magnetic flux \cite{kvas} $\Phi_{\textbf B_1} = \Phi_{\textbf B_2}$. This description involves secondarily applied equivalence with the internal flux $\Phi_{\textbf B_1}$ through the area delimited by the rotating disc. The existence of the external flux is omitted and the description is incorrectly simplified. As per this logic, the unshielded PHG should be functional as well, which is not true.}. In PHG, though, the internal flux lines move in the area of vectors $\textbf B_1$\footnote{Area of short flux lines between the magnet rings, which mainly emanate from the axial surface of pole $N$ and enter symmetrically into the counter-pole $S$ - area of the circularly homogenised magnetic field.} with respect to the internal standing wire $\vec i$. The external area of vectors $\textbf B_2$ in the vicinity of the external part of wire $\vec j$ with the most intensive magnetic field is shielded from the theoretical effects of the relative motion of flux lines. The shielding is over-dimensioned\footnote{The manufacturer declares a shielding efficiency of $100$ mT.} because the measured induction values of the external magnetic field amounted to an average of 23,76 [mT]. 
\begin{center}
\scalebox{0.48}{\includegraphics{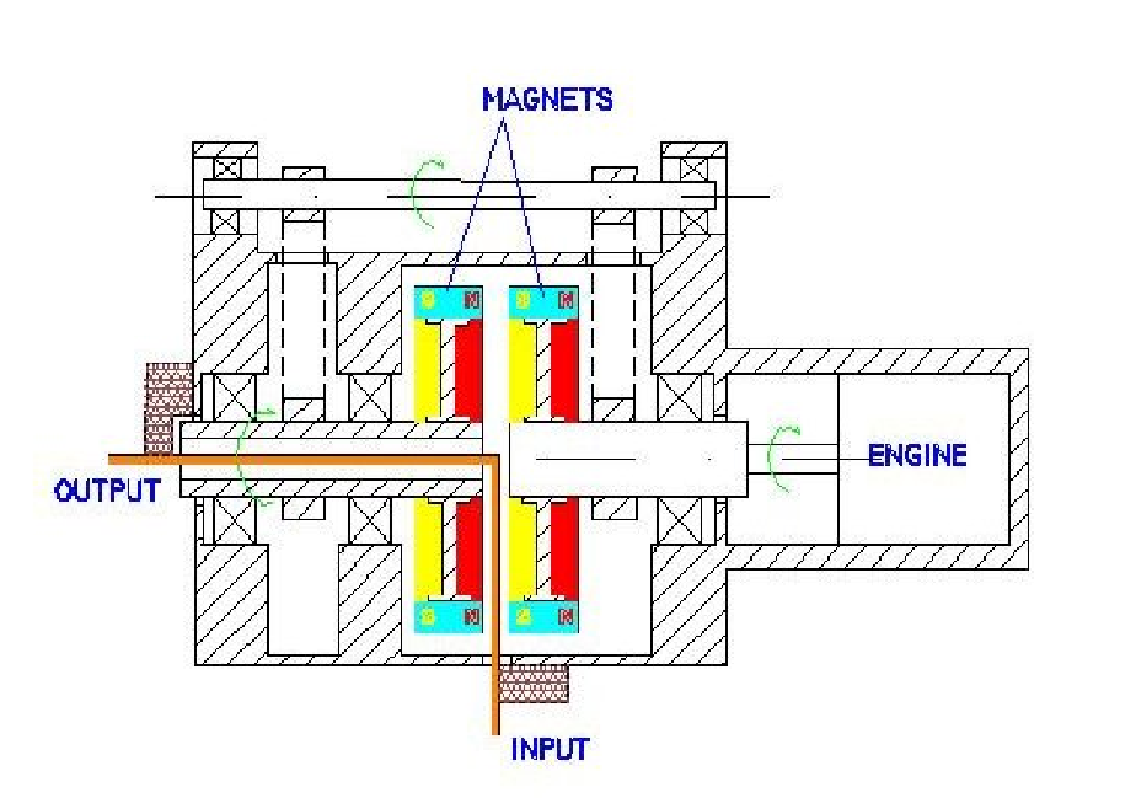}}\\
\captionof{figure}{\footnotesize{Schematic cross-section through technical solution of PHG, which allows the principal arrangement of two opposite and axially magnetised neodymium rings and the wire entering between these rotating ring and going out, as shown in Figure \ref{obr0}.}}
\label{obr1}
\end{center}
The maximum measured value was 96,6 [mT]. 
Due to the superconductive shield, it must occur that the magnetic field diverts at places of shielding and bypasses (the shortest of by way) the conductor without any influence on it.
For the circuit (coil) length $l$ in consideration, as shown in Fig. \ref{obr0} and in accordance with the chart in Fig. \ref{PHG-gr2}, the theoretical induction balance between positive and negative EMF contributions must be disturbed, i.e. $\mathcal E_m = \int\limits^l_0 \vec E_m \cdot \ d \vec l \not= 0$. There must be a surplus measurable induced voltage as a part of the circuit with a length of $j$ which does not contribute to the induction. Equivalence with the induction flux as per \eqref{U_m} is not possible as the shielding forms an open loop for the magnetic field. In terms of the occurrence of Lorentz force, the differential voltage induced in the shielded PHG must be equivalent to FHG. The blue line in Fig. \ref{PHG-gr2} shows the potential level which we should theoretically obtain from experimental data after shielding the return line $\vec j$.  It is this very experiment, which must determine the final decision as to the physical validity of the Maxwell equation \eqref{U_m}. 
\begin{center}
\scalebox{0.69}{\includegraphics{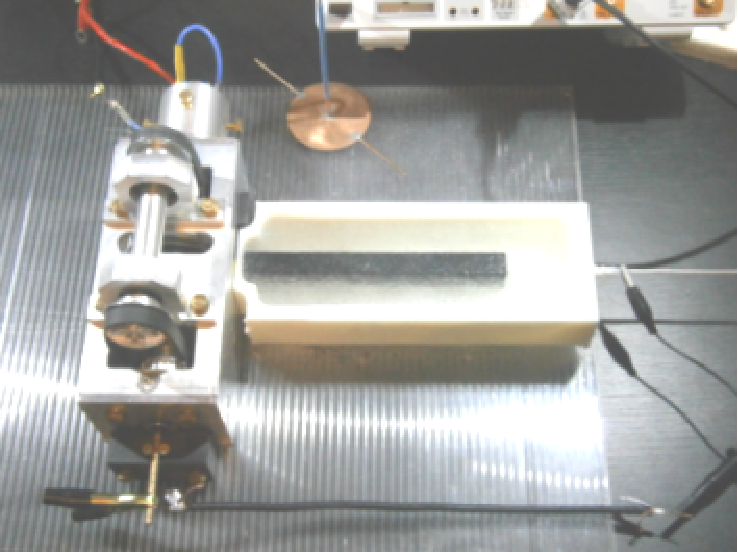}}
\captionof{figure}{\footnotesize{Brushless homopolar generator docked onto the superconductor shielding system in liquid nitrogen - real experimental model.}}
\label{obr1a}
\end{center}
\begin{center}
\scalebox{0.49}{\includegraphics{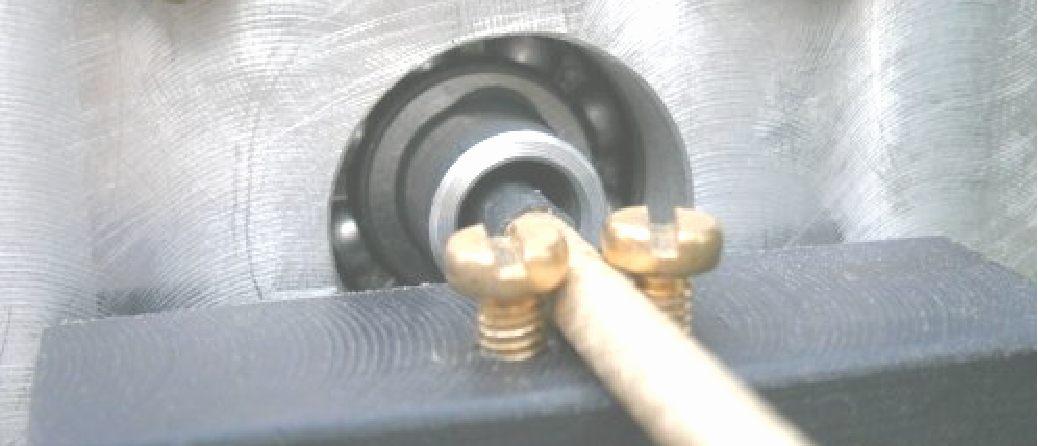}}
\captionof{figure}{\footnotesize{Detail of the conductive path with output through the hollow shaft.}}
\label{obr1b}
\end{center}    
Here we could elaborate in more detail on the topic of preserving continuity, which forms a basis for mathematical theorems used in electro-dynamics. The main assumption for using these theorems is the validity of the continuity equation \cite{kvas}, which preserves both the input and output flux. However, we know that the magnetic field actually does not represent any real flux.  It is rather an oriented field, which diminishes with growing distance from both poles. If we use a comparison to a flux and define a direction, it appears in such a way that the flux is diluted in the space and thickens again at the second pole\footnote{This comparison is more suitable for gases than liquids as in liquids, we speak of changes of velocity, not changes of density.} as if the field flew from the previously defined initial pole to the defined counter-pole without any losses. The dilution and thickening of the field flux should occur with a simple limitation that $div\vec B = 0$ in any measured area \cite{kvas}. If this limitation did not exist, then the field flux should decrease in the space with growing distance ($div\vec B < 0$) and increase again symmetrically to the original value when approaching the second pole ($div\vec B > 0$). In this case, in the unshielded PHG as shown in Fig. \ref{obr0}, a non-zero EMF should be theoretically induced with respect to the asymmetrical position of the wire loop $l$. The equivalence \eqref{U_m} would not be valid in such a case.  It is out of the scope of this paper to present a table of experimental values, namely due to the apparent remarkable coincidence with the assumption $div\vec B = 0 $.  The slight surplus $\mathcal E_m$ on the external loop shown in the chart in Fig. \ref{PHG-gr} is, in this case, rather caused by measurement error. 
                                                                              
We can conclude that the physical application of the continuity equation to a magnetic field actually only describes the theoretical course of flux lines as an analogy with liquid flow. However, this description does not imply, in logical terms, the ability of direct induction caused by mere movement of the wire in an idealised homogeneous field. It just corresponds, in mathematical terms, with the equivalence of the surface integral. For example, the analogy of the Biot–Savart Law, which describes the movement of an electrical charge in a homogeneous magnetic field, is also applied to an electrically neutral conductor - which is indeed not a direct logical method.  The only class of experiments, where the consequences of the movement of a conductor in an invariable magnetic field is objectively measurable, is based on PHG, which does not include any brushes. (FHG-based experiments are disputable, as the brushes can have a physical impact on the process.) 
It is the only experimental class which can objectively prove the validity of Maxwell's equations. The reason is simple: we are only able to simulate homogeneous radials of the magnetic field. All other experiments are based on heterogeneous fields. Theoretical accordance with practically measureable electro-dynamics is achieved using various electromagnetic constants e.g. in the form of environmental permeability.  This could lead to a false physical notion of induction occurrence.        

Tests on PHG with shielded measuring wire did not prove induction of any voltage at the level of one millivolt.
We have repeatedly used a shielding technology as follows: At first,  hypothermia of low temperature superconducting shielding was performed by means of liquid nitrogen at a distance of about $4$ [m] from the permanent magnets of PHG. PHG was subsequently docked to the shielding, and a millivolt oscilloscope was connected to it. When connected, the oscilloscope recorded closing of the circuit that was verified by an ohmmeter. After starting PHG up, the synchronous rotation of the two opposing magnets against the relatively standing unshielded wire began.  Synchronous movement of both axial magnets is achieved by the connection of toothed pulleys according to Figures \ref{obr1}, \ref{obr1a}. In order to en sure how is the induction behavior in the reference system that is moving in relation to the frame, at the end of the experiment, we repeated the conducted moving closer of the non-homogenous magnetic field of the neodymium magnet to the shielded part of the conductor $\vec j$, and to the non-shielded part of the conductor $\vec k$: induction did not occur at places with shielding, but induction did occurent at places without shielding. This demonstrates the ability and relevance of the usage of shielding. 
For comparison of the induced values between FHG and PHG, the PHG model was temporarily connected and tested as an FHG. In the FHG wiring, the induction amounted to 60 [mV]. Based on the data for the chart in Fig. \ref{PHG-gr2}, it was found  that the shielding in this PHG circuit eliminated $61$ \% of the external magnetic flux, which influences the external part of the wire loop. By its structural solution, FHG is able to eliminate up to $100$ \% of the overall internal magnetic flux (this can be achieved by the appropriate diameter of the copper disc). Comparison of this data shows that the PHG should induce the minimum value of $36$ [mV]. Since PHG does not have any brushes, this value should be considerably higher with respect to the low efficiency of FHG, due mainly to friction and heat losses on the brushes used. The overall course of the experiment is available on \cite{iv}.

The experiments carried out in this context have shown that to achieve induction in a magnetic field, it is necessary to meet, besides the known condition $\vec v \not\parallel \vec B$, another condition comprising also barely detectable situations itself: the instantaneous direction of motion of the conductor $\vec v$ is not collinear with some \emph {energy level created by vectors $\vec B$}. In other words, \emph {to cause the induction, the conductor must move across the energy levels.}

%\textcolor{blue}{
This energy level (hereinafter \emph{ESF}) in the radial distance $\vec r$ (e.g. with respect to the magnetic field's symmetry axis) can be  identified  with certain level (equipotential) levels of the magnetic field. For each pair of products of vectors $\vec B_{r_i}\vec r_{i}$, $\vec B_{r_j}\vec r_{j}$\footnote {We came to the product $\vec B . \vec r$ as follows: The level \emph{ESF} can have a defined normalized energetic permeability of size $W = \vec F_{max} . \vec r = Q \vec v \vec B . \vec r$, where $Q$ is a unit charge and $\vec v$ is a unit speed. Different energy levels may show identical induction levels $\vec B$ but distances $\vec r$ differing by the value of $\Delta_q$. If $\Delta_q = \|\vec r_{x_i} - \vec r_{x_j}\| > 0$ is the smallest possible difference of distance $\vec r$, then we term $\Delta_q$ the quantum difference.} with values of magnetic induction $\vec B_{r_i}, \vec B_{r_j}$ at points on closed curve $\vec s$, into which the position vectors $\vec r_i$, $\vec r_j$\cite{pach} point, we define the level \emph{ESF} as follows: $((\vec B_{r_i}\vec r_{i},\,\,\vec B_{r_j}\vec r_{j},\,\,\vec s) \Rightarrow (\vec B_{r_i}\vec r_{i} = \vec B_{r_j}\vec r_{j})) \Rightarrow ((\vec B_{r_i}\vec r_{i},\vec B_{r_j}\vec r_{j}) \in \emph{ESF})$. Two \emph{ESF} levels are energetically different if $(\vec B_{r_i}\vec r_{i} \in \emph{ESF}_i, \vec B_{r_j}\vec r_{j} \in \emph{ESF}_j) \Rightarrow ((\vec B_{r_i}\vec r_{i} \not= \vec B_{r_j}\vec r_{j}) \Leftrightarrow (\emph{ESF}_i \not= \emph{ESF}_j))$. On the contrary, two $\emph{ESF}$ levels are energetically equal if $(\vec B_{r_i}\vec r_{i} \in \emph{ESF}_i, \vec B_{r_j}\vec r_{j} \in\emph{ESF}_j) \Rightarrow ((\vec B_{r_i}\vec r_{i} = \vec B_{r_j}\vec r_{j}) \Leftrightarrow (\emph{ESF}_i = \emph{ESF}_j))$. Relationship of instantaneous velocity with tangent $\vec T$ of level \emph{ESF} is expressed by the equivalence $(\vec in \not\parallel \vec T) \Leftrightarrow (\vec in \not\parallel \emph{ESF})$. This analogy can be used for all other types of symbols $\parallel,\perp,\not\perp$. For the instantaneous velocity $\vec v$, at which the length difference $dl$ of the conductor is moving, we specify such referential component $\vec v_n$ for which is applied $\vec v_n = \cos(\alpha)\vec v$ and $\vec v_n \perp \emph{ESF}$ (level \emph{ESF} is usually formed by closed curves or the cylindrical surfaces). Further we define:  
\begin {eqnarray} \label{oI}  
  \text{If $dl$ passes through two different \emph {ESF} at speed $\vec v_n$}\nonumber \\
  \text{then there exists a speed $\vec v_\perp$ which is referred to as}\nonumber \\
  \text{\emph{transverse speed}}.
\end {eqnarray}
If there is $\vec v_\perp(\alpha,\vec v)$ then $\vec v_\perp \equiv \vec v_n$, and it is thus perpendicular to the energy levels \emph{ESF}, and therefore meets the new condition of reaching induction.

Furthermore, for simplicity we state $\vec v = \vec v_\perp$. The magnitude of the induced voltage will therefore be proportional to the velocity $\vec v$ of the conductor and to the magnitude of the magnetic field density. (Faraday's Law from the point of view of line integral in \eqref{U_m} - in terms of Lorentz force). We determine the density of the magnetic field from the measurement as the ratio of the force to the electric charge velocity $\vec F_{max}/Q.\vec v = \vec B$. The induction capability is the function of the resistance of the conductor’s passability through the magnetic field. In the perpendicular direction $\vec v$ to the energy level \emph{ESF} is its passability, accompanied by the greatest counteraction and thereby the greatest ability to generate an induction.
\begin{center}
\scalebox{0.34}{\includegraphics{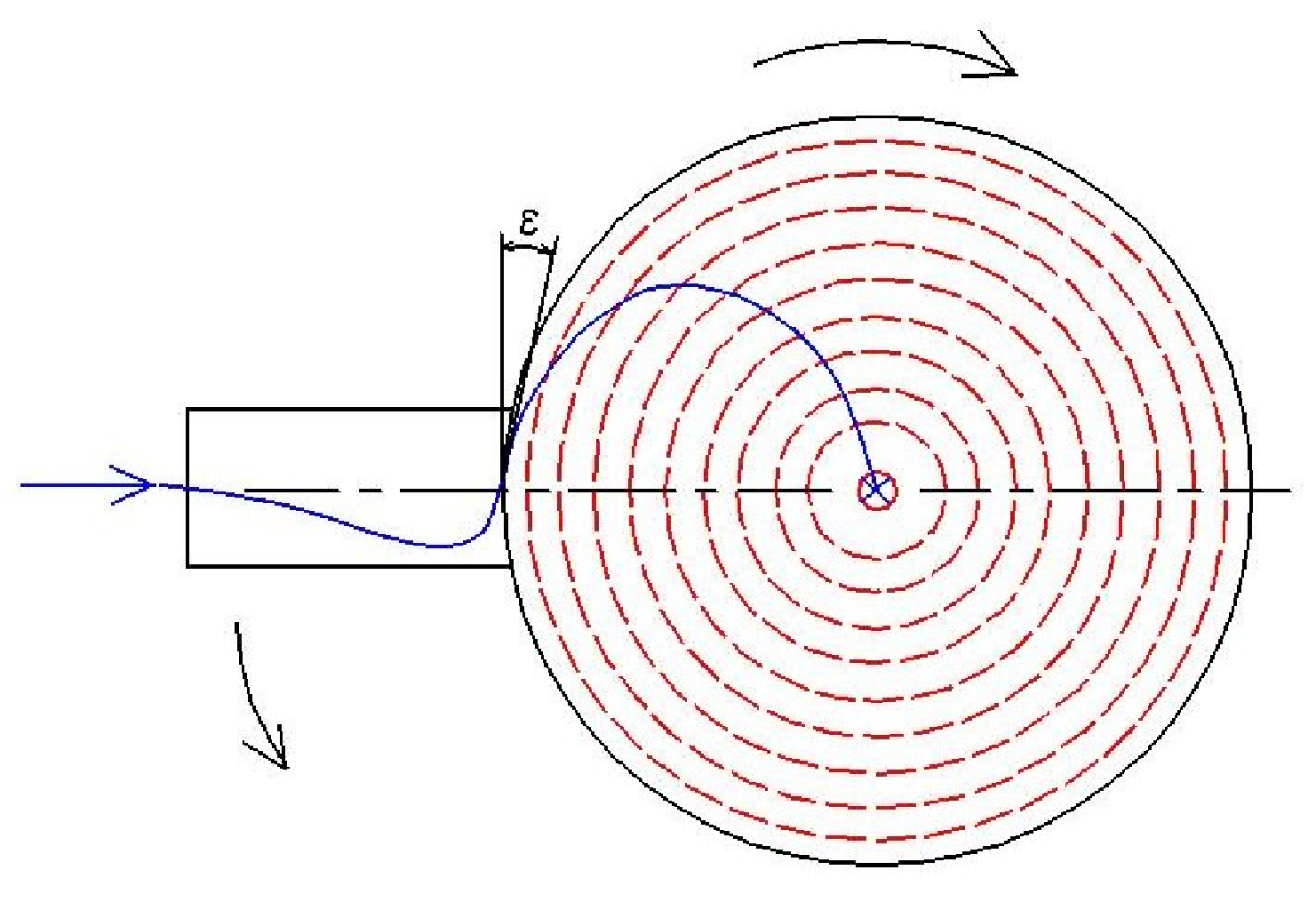}}
\captionof{figure}{\footnotesize{The notion of the middle track of the conductivity path in the rotating FHG.}}
\label{obr2b}
\end{center}
\hspace{-1cm}
\begin{center}
\scalebox{0.56}{\includegraphics{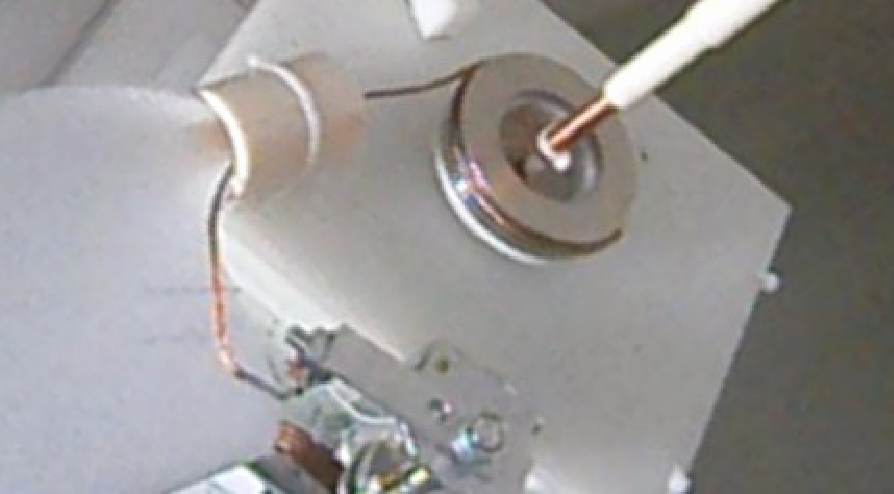}}
\captionof{figure}{\footnotesize{The experimental model with the flexible conductor, which is winding up on the axially magnetised magnetic disc.}}
\label{obr2a}
\end{center}
One of the experiments that demonstrates the behaviour of the flexible conductor and can serve as an alternative explanation of the induction in the FHG is seen in the Figure \ref{obr2a}. Winding/unwinding of the flexible conductor against a peripheral wall of the rotor of Faraday’s disc causes the DC induction of the voltage there in the conductor \footnote {The automatic winding of the flexible conductive cord takes place also inversely when the conductor is under electric current.}. Brushes of FHG basically represent simulation of the internal winding of the conductor to the periphery of the magnetic disk and thereby simulate the continuous passing of the conductive portion of the disk volume through the \emph{ESF} in the direction of the transverse velocity $\vec v_\perp$, as it is schematically shown by the blue curve in figure \ref{obr2b}.%%%%XXXXB 
In this case, by connecting the load, we connect the circuit with the FHG centre. Thus the shortest conductivity path with the lowest potential will be created. During disc rotation, due to the relative movement of the brush versus the rotor surface, this path with the potential pit (channel) starts prolong off and in contact surface is inclined at the angle $\pi/2 - \varepsilon$.

This is followed by counter-action in accordance with Lenz Law \footnote
{The original law does not specify what kind of change it is - the change of magnetic flux was attributed to it later.}\cite{lenz}:
The increase of the electric intensity along the potential pit generates a current. Subsequently free electrons from the vicinity are captured into the potential pit.
%%%%XXXXE
These electrons accumulate in a wave that is synchronous with the relative movement of the brushes. In such an arisen electron wave, there exists a steady excess of charge that can be detected by measurement. Lorentz force\cite{lore} arises only by moving the conductor transversely over levels \emph{ESF}.

The more than 150-year-old concept of magnetic flux, or the concept of induction lines, creates a false illusion of the immanence of magnetism in the Euclidean space and obscures the real dynamic properties of magnetically influenced physical space. The experiment built on homogeneous radials in the sense of isomagnetic levels of \emph{ESF}, as in the Figure \ref{obr1a}, is unique, original and allows a special case to be ensured when $\vec v \parallel \emph{ESF}$ and at the same time $\vec v \perp \vec B$ occurs. Experiments based on Helmholtz’s coils or on the basis of improved Maxwell’s coils and also most technical solutions achieve the optimal setting $\vec v_\perp$. This is caused by the function of the benefits of using different technical symmetries. It is then difficult to reveal another general property of the magnetic field, when no induction occurs: namely, if $\vec v$ is collinear with some level \emph{ESF}, then $\vec v \parallel \emph{ESF} \Rightarrow \vec E = 0$. The layer \ emph {ESF} may also include a special situation of levels of direct current conductor, when the inductor vectors are collinear with velocity vector $\vec v \parallel \vec B \Rightarrow \vec E = 0$ and at the same time $\vec B \parallel \emph{ESF}$. On energy levels of the straight current conductor, we can demonstrate another special case, when $\vec v \not\parallel \vec B$ and at the same time for the considered circular energy level  $\vec v \not\parallel \emph{ESF}_1$ occurs, and yet induction does not occur. This is because in the vicinity there is no energetically different circular level $\emph{ESF}_2$, and then no transverse velocity $\vec v_\perp$ exists. Both levels $\emph{ESF}_1$, $\emph{ESF}_2$ belong to the common level \emph{ESF} of the magnetic field on the cylindrical surface. As a conclusion from this paragraph, we can state that the energy levels of magnetic field vary dynamically depending on the layout of these fields.
\\

{\bfseries Transformation of Faraday's Law from idealism to approach to realism}
\vspace{0,1cm}

In differential form, Maxwell's interpretation of Faraday’s experiment is defined as follows (Faraday’s Law)\footnote{In this differential form of Faraday’s Law, note that in the homogeneous magnetic field, the absolute value of $\vec B$ does not have to change. At the stationary position $\vec B$, this fact would in \eqref{rotE_m} imply zero value $rot\vec B \Rightarrow d\vec B = 0$, and hereby zero contribution to electromagnetic induction. However, partial changes of vector $\vec B$ ensure, according to the convention with liquid flow (Stokes’ Theorem), transport’s and rotational changes of vector coordinates $\vec B$. These hypothetical and unverifiable changes of coordinates $\vec B$ ensure the theoretical usability of Maxwell's equations even for a non-existing homogeneous magnetic field.}:
\begin {eqnarray}  \label{rotE_m}
   rot\vec E_m = - \frac{\partial \vec B}{\partial t} \Longleftrightarrow rot \vec E_m = rot(\vec B \times \frac {\partial \vec x}{\partial t})  
\end {eqnarray}

Changing the induction flow implies induction only by moving the conductor in the general magnetic field, and thus there exist some parts of the moving conductor, for which the velocity is $\vec v \not\parallel \vec B$. It is a very general condition, and it is weak for a real magnetic field. It is a criterion too general and too weak for a real magnetic field. It does not correspond to reality, as was revealed from the behaviour of shielded PHG. To this criterion we therefore add, according to \eqref{oI}, a narrowing criterion $\vec v \not \parallel ESF$. Complete formal definitions of both criteria follow in the next paragaph.

\begin{itemize}
  \item In the general relation $\vec E_m = \vec B \times \vec v$ for a wire moving in a magnetic field, $\vec{v} \not{\parallel}\,\, \vec B$ must hold.
  \item In the general relation $\vec E_m = \vec B \times \vec v$ for a wire moving  in a magnetic field, the wire moves across different energetic levels \emph{ESF}. Thus, $\vec v_\perp$ exists, and so $\vec v \not\parallel \emph{ESF}$ applies. 
\end{itemize}
In accordance with these terms, and using the definition \eqref{oI}, Faraday's Law inevitably narrows from the point of view of Lorentz force (from the point of view of the line integral in \eqref{U_m}):
\begin {eqnarray} \label{I}
  \vec E = \vec B \times \frac{\partial \vec x_{\perp}}{\partial t},
\end {eqnarray}

where $E$ represents instantaneous induced intensity in a conductor of length $dl$ moving at the speed $\partial \vec x_\perp / \partial t$, where the perpendicular mark $\perp$ means that for velocity $\vec v = \partial x/\partial t$, there exists some moving component $\vec v_\perp = \partial \vec x_\perp / \partial t \not= 0$ perpendicular to \emph{ESF} levels. The statement $\partial\vec x_\perp$ represents the relatively shortest path between \emph {ESF} levels, which the difference $dl$ of the conductor travels. The relationship indicates that the only physical reason for the rise of the induction is movement of the perpendicular velocity component to the \emph{ESF} level at $\vec v \not\parallel \vec B$. Thus, besides the magnitude and direction of the induction vector, the magnitude and direction of the velocity vector, we are also interested in the magnitude of the deflection of the velocity vector from the normal line of the \emph{ESF} level. For the interval $||\vec v_\perp(x_i)|| \le ||\vec v(x_i))||$  we can intuitively define instantaneous usability $\gamma(i)$ of the \emph{ESF} level in the point $x_i$  (rectangular coordinates in 3D, $x_i \equiv x_{i_1},x_{i_2},x_{i_3}$) using the velocity vector $\vec v(x_i)$ and its component $\vec v_\perp(x_i)$ perpendicular to the \emph{ESF} level: 
\begin {eqnarray}    \label{III}
\gamma(i) = \vec v_\perp(x_i)/\vec v(x_i) = \cos(\alpha),
\end {eqnarray}
where $ x_i $ is the point to which the spatial radius $\vec r_i$ (spherical coordinates) leads as the argument of the vector function of induction $\vec B(\vec r_i) = \vec B$\,\,\cite{pach}.  
$\alpha$ is the angle between direction of the velocity $\vec v$ and its motion component $\vec v_\perp$ in the direction of the relative path $\partial\vec x_\perp$. By putting \eqref{III} into 
\eqref{I}, we immediately get: 
\begin {eqnarray}   \label{E1}
  \vec E = \vec B \times \frac{\cos(\alpha)\vec v.\partial t}{\partial t} = \cos(\alpha)(\vec B \times \vec v).
\end {eqnarray}

In the case of integration of \eqref{rotE_m} over the surface, we get the relation \eqref{U_m}, and in the case of integration \eqref{E1} over the length of the conductor, we get a formally similar equation for EMF:
\\
\begin {eqnarray}   \label{U_e}
\mathcal E = \oint\limits_l \vec E \cdot d\vec l       
\end {eqnarray}
\\
In terms of the considered conductor, $\vec E$ is oriented as $\vec E_m$ but with different values. Besides the angle between $\vec B$ and $\vec v$, values of $\vec E$ are influenced by $\cos(\alpha)$.

By the product of \eqref{III} with $\vec E_m = \vec B \times \vec v$, we get a relation for instantaneous induction, that expresses the ratio between Maxwell's concept of electric intensity $\vec E_m$ 
and the concept in the context of the experiment described here:
\begin {eqnarray} \label{E2}
\vec E = \gamma(i) (\vec B \times \vec v) = \vec B \times \frac{\vec v_\perp(x_i).\vec v}{\vec v(x_i)} = \nonumber
\\
\vec B \times \vec v_\perp(x_i) = \cos(\alpha)(\vec B \times \vec v)          
\end {eqnarray}
We get an equivalence with \eqref{E1} immediately. It should be mentioned that the general equation \eqref{E2} or an identical equation \eqref{E1} represents just a relationship between $\vec E_m $ and $\vec E $. Only interactions according to \eqref{I} are physically relevant,  i.e. only the motion of the conductor across levels \emph{ESF} at $\vec v \not \parallel \vec B$.

For simplicity, for a moving conductor in a non-homogeneous magnetic field 
(by analogy to the figure \ref{obrz} below) in the whole integrative range of the field $\vec S$, we introduce a coefficient $ \gamma_\phi = (\int\limits_S \gamma(i)\cdot d \vec S)/ \vec S $ as an \emph{average usability of the ESF levels} of the magnetic field. We can the write the equation
\begin {eqnarray}  \label{fE_m}
\gamma_\phi\int\limits_S \, rot\vec E_m(\vec x_i)\cdot d\vec S \,\, = \oint\limits_l  \vec E(\vec x_i)\cdot \ d \vec l \,\,.
\end {eqnarray}
If we integrate both sides of \eqref{fE_m} (lhs over the area bounded by the conductor, rhs over the length of the conductor), we get the scalar result:
\begin {eqnarray}   \label{fU_m}
 \gamma_\phi \,\,\, 
  \mathcal E_m =  \mathcal E .
\end {eqnarray}

This result illustrates that we can obtain induced voltage only by moving the conductor to another level of \emph{ESF}, by which the energy is directionally polarised and preserved. It is also the path of the greatest resistance to passing between the levels. The effects of these levels are demonstrated by the analog (continuously) in relation to the goniometric function. There is a noticeable correlation with the energy levels of electrons. In the macroscopic world, these levels are not discretely (discontinuously) determined. It is important for the moving conductor to know what degree of freedom it has towards the layer \emph{ESF}. It is evident that the vector of induction $\vec B$ is not a simple vector, as is perceived in Maxwell's vision. It is completely dependent on the ambient vector topology, and this topology plays a major role in whether an induction occurs or not.
 
It follows from the nature of the function $\cos$ that for about $67 \%$ of technically executable experiments (angle $\alpha$ between $\vec v$ and $\vec v_\perp$ achieves up to $60^\circ$ from a maximum $90^\circ$) the result will be in accordance with Faraday's Law, with an error up to $50 \% \leftrightarrow \cos(60^\circ)=0.5$ from predicted values. Thus, Stokes’ Theorem is usable in at least $67 \%$ of all possible applications. Thus, by the \eqref{fU_m}, Stokes’ Theorem is not applicable for physically theoretical considerations, aside from the continuity preservation. The interaction of the moving conductor with the magnetic field becomes our main interest.

If for the majority of known experiments and technical solutions, there is a mean deflection $\alpha \ll 45^\circ$ of the velocity $\vec v$ from normalcy \emph{ESF} according to \eqref{I}, \eqref{E1} and \eqref{E2}, then there must be average usability $\gamma_\phi \approx 1$. The predictive ability of relations \eqref {U_m} and \eqref {rotE_m} has an error of up to $30 \% \leftrightarrow \cos(45^\circ)\approx 0,7$ in the worst case. 
For values $\gamma_\phi$ in the interval $0 < \gamma_\phi \ll 0,7$ the new relation gives a necessary correction of Faraday’s Law. For values $\gamma_\phi = 0$, the result will correctly predict the behaviour of shielded PHG in accordance with \eqref{fU_m}. Thus, there is no doubt that in practice, the occurrence of average usability of \emph{ESF} levels can be estimated in the interval $0,5 \ll \gamma_\phi < 1 \leftrightarrow  60^\circ \gg \alpha > 0$.
\\                                         
\section*{Discussion}    

According to the above experiment, relations \eqref{fE_m}, \eqref{fU_m} predict that moving the electrically neutral conductor in the homogeneous magnetic field, as shown in the picture \ref{obrz} below, will be $\mathcal E = 0$. This is due to the fact that, in every direction of the velocity $\vec v$, there is a collinear level \emph{ESF} and the condition $\vec v \not\parallel \emph{ESF}$ can never be met. So it is not possible to obtain energy by changing energy levels. According to \eqref{I}, any Lorentz force will not be produced, therefore no voltage will be induced because $(\vec v_\perp = 0) \Rightarrow (\vec E = 0)$. In terms of induction, the relationship \eqref{rotE_m}  in this context is considered to be physically inaccurate, creating a hypothetical and unverifiable idea of magnetic field turbulence. It only describes time changes of geometric quantities in 3D space. The equations \eqref{E2} and \eqref{fE_m} show the necessary theoretical correction of the current equations so that they can predict result of the experiment described here.

\begin{figure*}[ht]
\centering
\scalebox{0.45}{\includegraphics{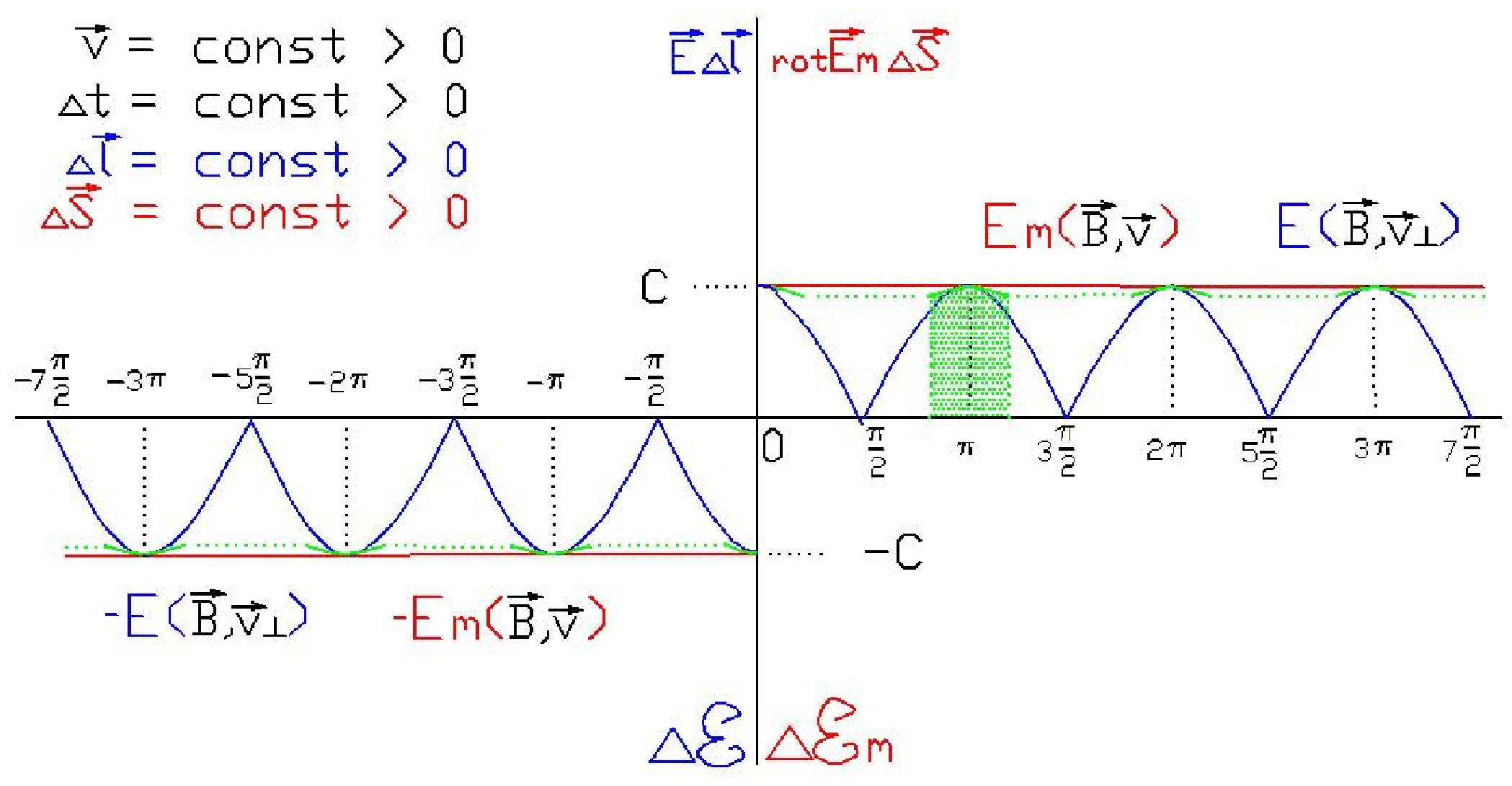}}
\caption{\footnotesize{
Graphic comparative study of Maxwell's regulation $E_m$ and the course of the herein proposed concept $E$ dependent on the angle $\alpha$ with a common non-zero vector $\vec v$.}}
\label{graf2}
\end{figure*}

If we set in accordance with the result \eqref{fU_m} $\vec B = const(\vec B) > 0$, $\vec v = const(\vec v) > 0$ and firmly set  non-zero increments of the length $\Delta\vec l$ of the conductor, we get two prescriptions in the image \ref{graf2}: the graph of the dependence of the EMF creation on the medium deflection of the velocity $\vec v$ from the component $\vec v_\perp(\alpha,\vec v) = \cos(\alpha).\vec v$ according to the prescription $E(\vec B,\vec v_\perp) = \mathcal E$ (blue curve), and the graph of the dependence of the EMF creation according to the prescription $E_m(\vec B,\vec v) = \mathcal E_m$ (red line). Changing the deflection of the movement of the differential lengths $dl$ of the conductor from the normalcy level \emph{ESF} causes in concept $E(\vec B,\vec v_\perp(\alpha,\vec v))$ the change $\mathcal E$, but in the concept $E_m(\vec B,\vec v)$ it stays $\mathcal E_m = C$. 
\\

It can be seen from the Figure that at $\vec v = const$, the Maxwell's ${\mathcal E}_m = E_m(\vec B,\vec v)$ is independent of the  deflection angle $\alpha$.  $\alpha = n\pi/2$, for integers $n$. Thus it assumes nonzero values ${\mathcal E}_m$ also for values $\alpha = n\pi/2$, for whole $n$. If we consider the presented suggestion with the prescription $E(\vec B,\vec v_\perp)$ as the best match to physical reality, and at the same time we promote Maxwell’s concept as the best proven in practice then, for example in the interval of values of the angle $0 < \alpha \ll \pi/4 \Rightarrow 1 > \gamma_\phi \gg 0,7$, differences are in no way dramatic. This is ensured by the course of the function $\cos(\alpha)$ that even with large scattering of  values $\alpha = n \pi (\pm\pi/4)$ gives a  decrease $\mathcal E$ of less than $30\%$. For less scattering, values $\mathcal E$  are asymptotic with the course $\mathcal E_m$ according to the relation \eqref{fU_m}. Actual measured values (green arcs) can be even more asymptotic with red lines, when considering the formation of electromagnetic constants according to the concept $E_m(\vec B,\vec v)$. In the methodology of their determination, the average usability of the level \emph{ESF} is being covertly included. The green hatched section and sections under the green curves represent technically easily verifiable areas with high usability of levels \emph{ESF}. The green dotted segment represents the theoretical assumption for creation of EMF according to the concept $E_m(\vec B,\vec v)$. This is technically an inferiorly accessible area. Our experiment concerns this area and shows that current theoretical assumptions are erroneous. %%%% end
\\             
There are three advantages of the presented concept:
\begin{itemize}
  \item It implies the prediction of the PHG behaviour and thus respects the results of a wider range of experiments.
  \item It can give equivalent results consistent with previous practice, because the average values $\gamma_\phi$ is related to the non-closed area ~ $\vec S$ likewise the magnetic flux. 
  \item It reveals the phenomenon of the polarised transmittance of the magnetically influenced space and thus opens up a wide range of formerly hidden possibilities of new technical realisations.
\end{itemize}              

This article describes a pure homopolar generator. The same formula applies to the inverse phenomenon in the pure homopolar motor PHM\footnote{Meant as Pure Homopolar Motor: i.e. a motor that does not need any brushes, electronics or semiconductor and is powered directly by DC current.}        

Examples when theory does not provide a completely credible theoretical basis for engineering practice are, for example, the grant of the patent \cite{USi,CZi,AUi}. This group of patents can not be functional even from the point of view of Maxwell's electrodynamics and in this case contradict the law of the conservation of continuity, hence the energy conservation. Furthermore, patents \cite{USk,DEk,CNk}, which should be theoretically functional similar to PHG/PHM. These patents assume the creation of an imbalance of EMF in a continuous and homogenised magnetic field using high temperature shielding components. As significant support for finding discrepancy between theory and practice, we consider the fact that, despite the advantages of similar solutions, none appears in practice. For example, there is no wind power plant that could use similar solutions to utilise and directly  produce  DC current in the sense of PHG. Classic high performance commutating (brushed) DC generators are commonly used. Brushless AC generators with electronic rectification for subsequent further processing into the distribution network have come to the forefront in recent years.

The intention of this report is to point out, within the deeply rooted myth of contemporary theoretical electrodynamics, that there could be a technical solution in the sense of PHG/PHM, which would use some theoretical homogeneous or actual homogenised magnetic field for the creation of inductive current/torque. If a model of a design is not created made, the creator may never learn of the mistake. There will be more reasons why feedback from industry to academic awareness is blocked and we'd rather not speculate about them.

\section*{Methods}               

The paper clearly shows that we used the oldest physical methods, i.e. methods of model construction for a functional demonstration. We used our own idea of identical simulation of the theoretical assumptions of the function of FHG on PHG. Finally, we used the current method of shielding with high-temperature YBaCuO superconductors, whose efficiency - as declared by the manufacturer - exceeds $100$ [mT]. To maintain the appropriate temperature, the standard cryogenic technology based on liquid nitrogen was used. The measurement itself did not require any special equipment or methods. For the sake of simplicity, we compared the induction values on FHG with dimensions and parameters identical to PHG. The resulting values on FHG exceeded $10$ [mV] by far. The measurement on PHG was carried out using millivolt oscilloscopes HMO722 or HMO2008 with a resolution of $1$ [mV]/Div (HMO722 has $1$ Div=50 px, diagonal $16.5$ [cm] $\Leftrightarrow$ 8x12 Div) with probes HZ154 or HZ200 and a 3D teslameter Helimag MP$-1$ with complementary software for processing of the measured values. The chart was created using the previous version of the physical SW. We actively investigated what was happening in the demonstration devices down to the level of [mV]. Further specification was pointless,  due to the noise amounting to $\pm$0.25 Div at these low voltage values. The result of this experiment is significant. In the analysis, we  used the description variant that preserves the vector nature. It is easy to see that this analysis continues what is called Faraday’s Law. Furthermore, at the boundary point, it explains the cause of its failure and corresponds with reality.

%\def\bibname{}                                                                        
%\bibliography{ref}

\section*{Acknowledgments}

We would like to thank CAN SUPERCONDUCTORS, Ltd. for their helpful approach in the implementation of our experiment, reducing the price of high temperature superconducting shielding 100 [mT] by half. Our thanks go particularly to Dr. Vladimir Plechacek and Dr. Jan Plechacek for their professional technological and organisational realisation of the production of this custom superconducting shield. 
%A big thank you for his explanation of FHG also goes to Prof. Mgr. Tomas Tyc, PhD, of the Institute of %Theoretical Physics and Astrophysics - Physics Section - a Faculty of Masaryk University, Brno.

\section*{Author contributions statement}

P.I. designed the study, performed the experiment, analysis and wrote the paper. M.I. performed the final proofreading and an English translation from the perspective of mathematics. Both authors reviewed the manuscript.

\section*{Additional information}

Corresponding author: P.I. Competing financial interests: The authors declare no competing financial interests.

%\begin{center}
\newpage

\end {multicols}

\end{document}